# Vulnerability Modeling for the Adaptation of Physical Systems to Climate Change Extremes

Matthieu Dutel

---

Climate change is modifying the frequency and intensity of extreme hazards in a non-uniform spatial manner, increasing the risk of damage for existing physical systems. Vulnerability modeling based on vulnerability curves provides a quantitative link between hazard intensity and physical damage, supporting adaptation strategies. This paper proposes conceptual frameworks for vulnerability models of physical systems in the context of climate change, aimed at supporting vulnerability-based adaptation. A theoretical use case illustrates the methodology: potential damage on power transmission towers facing wind gust hazards across France, modeled through vulnerability curves. Despite strong assumptions, this theoretical case study demonstrates how vulnerability models can guide adaptation by reducing intrinsic vulnerability, adapting the physical system to maintain acceptable damage levels under evolving climate extremes.

## 1. Introduction

**List of Abbreviations**

| **Abbreviation** | **Signification** |
|---|---|
| PDR | Physical Damage Rate |

Climate change is fundamentally changing the statistical distribution of climate extremes, leading to non-stationary hazard intensities that increasingly exceed historical design assumptions or imply damage for many physical systems and infrastructures. A broad scientific consensus has emerged on the change in intensity for a given return period, the spatial change, the change in frequency for given extreme climate events. These extreme events include extreme winds, extreme temperatures, extreme precipitation, droughts, floods and compound events. These changes expose infrastructures to climate intensities outside the design hypotheses. Many of these infrastructures were designed under stationary climate assumptions. As a result, existing infrastructures face growing risks of physical damage and or service disruptions (i.e failure) over their operational lifetime.

In this context, modeling quantitative physical vulnerability by linking hazard intensity and the physical damage rate experienced by a physical system is important to preserve performance of physical systems. Vulnerability modeling based on vulnerability curves provides the quantitative link between climate intensity and physical damage rate for a given system. Physical vulnerability modeling enables infrastructure managers, engineers, and public authorities to anticipate how infrastructure adapt to potential future climate extreme intensities. The vulnerability curves directly support adaptation decisions because they quantify the physical damage rate expected under changing climate conditions for a given system. Their role is distinct from fragility curves, which express the probability of failure rather than expected physical damage rate. While fragility information is crucial for network-level resilience studies on a functional part, physical-vulnerability curves seem to be a particularly well-suited choice for planning physical adaptation actions on assets and subsystems.
Most existing infrastructures were not designed to face non-stationary climate extremes. At a given location, if hazard intensity increases, the vulnerability of a given system, if unchanged, implies increasing physical damage rate as the intensity increases. Maintaining acceptable damage levels therefore requires adaptation actions, either by reducing exposure or by reducing intrinsic vulnerability in the way that, for a given hazard intensity, the physical damage rate is reduced. This can be conceptualized as a rightward shift of the vulnerability curve on the intensity axis: the system becomes less vulnerable to the same intensity threshold. Effective adaptation thus requires reliable physical vulnerability modeling that relates climate intensities, system characteristics, and expected damage.
Despite the conceptual clarity of vulnerability modeling, existing vulnerability information remains heterogeneous, and still incomplete; i.e., there is no existence of a vulnerability curve for each system type, each system level, each hazard. The availability of vulnerability curves differs between hazard types, system types, structural designs, regions, and system levels. In many cases, vulnerability curves are unavailable, inconsistent with local design practice. Also, information to build them can be proprietary. The lack of standardization of vulnerability curves across systems and hazards limits the capacity to compare vulnerabilities in a consistent manner, reduces the ability to perform long-term adaptation planning, and prevents direct quantification of expected physical damage rate under different scenarios of evolving climate extremes. To contribute to addressing this gap, this paper focuses on the formulation and application of a generalizable vulnerability modeling framework for physical systems under climate-related hazards. The objective is not to produce a universal vulnerability curve – which is neither feasible nor desirable – but to propose a systematic formulation that can incorporate diverse vulnerability information into a coherent model. This vulnerability model would link climate intensity, system characteristics, and physical damage rate. This method enables the computation of intensity-based physical damage rate for individual assets or ensemble of assets and provides a consistent basis for vulnerability-based adaptation planning. To illustrate the methodology, the paper presents a theoretical use case involving power transmission towers facing wind gust across France. Although the case relies on many very strong assumptions, particularly regarding system characteristics and the transferability of vulnerability curves, it demonstrates how the proposed modeling approach can be applied in practice, how damage projections can be produced under given return-period intensities, and how adaptation actions can be derived from vulnerability information to maintain acceptable damage levels in a changing climate.

The remainder of this paper is structured as follows. Section 2 presents the research gap that motivates the need for general vulnerability models and vulnerability-based adaptation of physical systems to climate change extreme intensities. Section 3 introduces the core concepts and definitions, including the distinction between vulnerability curves and fragility curves. Section 4 formulates the proposed vulnerability model. Section 5 presents the methodology for its use in adaptation. Section 6 presents the theoretical use case, including the modeling setup and results. Section 7 discusses the limitations of the approach. Section 8 concludes with implications for future research on vulnerability-based adaptation of physical systems.

# 2. Research Gap

### 2.1. Climate Change, Non-Stationarity, and the Need for Quantitative Vulnerability Modeling

Climate change is intensifying, redistributing spatially, and locally amplifying extreme climate intensities in a fundamentally non-stationary manner. The frequency, spatial distribution, and amplitude of extreme temperatures, intense precipitation, extreme winds, droughts, storms, floods and compound events are changing. As highlighted by major assessments and scientific reviews, physical systems and infrastructures will increasingly face hazard intensities that exceed the conditions for which they were originally designed. Existing infrastructures were designed under assumptions of stationary extreme climate inputs. As climate extremes evolve, systems are increasingly exposed to intensities that surpass design thresholds, implying an increasing risk of physical damage and functional damage over their operational lifetimes.

This evolving context, documented by international assessments (e.g. IPCC AR6), sectoral reviews, generates a need for models that quantitatively link climate intensity to physical system response. Quantitative vulnerability information is essential for design, for maintenance management, and for reliability of physical systems. In adaptation planning, such quantitative relationships, particularly between climate intensity and physical damage rate, constitute indispensable information for anticipating how physical systems will be affected by future climate extremes, determining acceptable damage levels, and guiding adaptation strategies.

Within this perspective, physical vulnerability curves (also known as vulnerability functions or damage functions) represent a central tool. These curves associate a hazard intensity with an expected physical damage rate. These curves can be used at the asset scale and can be aggregated over many assets; they can also be curves which a non-specific to one system type. These curves provide direct, intensity-conditioned information necessary for physical adaptation planning. By contrast, fragility curves are linking a climate intensity to a probability of failure. Fragility curves are key for network level resilience studies but seem to be less suited than vulnerability curves for providing physical damage rate required for quantitative, damage-based adaptation decisions. Vulnerability indicators, while providing detailed system-specific and location-specific information, do not encode explicit intensity-damage relationships and therefore do not provide precise physical damage rate values.

Thus, although indicator-based methods are complementary for capturing system-specific or subsystem-specific intrinsic vulnerability, they are not the most suitable tools for large-scale, intensity-based adaptation decisions. They are nevertheless widely used. When the resources to construct a vulnerability curve (time; intensity-based system damage rate) are not available, vulnerability indicators are an interesting method.

### 2.2. Heterogeneity of Existing Vulnerability Models

Despite their conceptual relevance, vulnerability curves are far from available for all assets, hazards, structural designs, system levels, or regions. Existing vulnerability information remains heterogeneous, fragmented, and frequently incomplete. Studies across hazards and systems -such as flood-related vulnerability curves for buildings (Maria Papathoma-Köhle et al., 2022), agricultural vulnerability curves (Monteleone et al., 2023), wind-related vulnerability curves for transmission towers (Reinoso et al., 2020), or debris-flow models (Zhang et al., 2018) show a great diversity in which the vulnerability model is applied (for different systems, for different hazards). Additional work shows the heterogeneity and also the broad scope of vulnerability curves : vulnerability curves differ across countries (Huizinga et al., 2017). Vulnerability curves exist for transport assets (Habermann and Hedel, 2018). These examples illustrate the large scope of vulnerability models.
The heterogeneity of vulnerability curves in different studies, and the fact that there is not one big vulnerability model containing all the vulnerability information and thus able to give for each intensity of each hazard a physical damage rate comes from the fact that producing even a single model for one hazard, one system, and one geographical context requires extensive data, expert knowledge, and methodological choices. As seen in the Papathoma-Köhle and colleagues ensemble of works (Maria Papathoma-Köhle et al., 2022; M. Papathoma-Köhle et al., 2022; Papathoma-Köhle, 2016; Papathoma-Köhle et al., 2019) , developing a vulnerability model – whether vulnerability curves or vulnerability indicators – for a specific hazard-system-location combination demands significant effort. Consequently, studies tend to fall into separate categories: those that build or refine a specific vulnerability model; those that collect existing models; and those that apply vulnerability models for strategic decision-making or adaptation. Because performing all these tasks at once is extremely demanding, existing models accumulate in a set of vulnerability curve manner rather than forming a unified and general vulnerability model.

### 2.3. Limited Availability of Vulnerability Curves across Systems, Hazards, Designs and Regions

Beyond their heterogeneity, vulnerability curves are also limited in availability: many system-hazard pairs lack a well-suited vulnerability curve. Relevant vulnerability information may exist but remain proprietary or inaccessible; in other cases, for hazards related to climate change, the system at a given location may never have faced the extreme intensities projected for the future, so that the corresponding damage information simply does not exist.

Even where vulnerability curves are available, they are often design-specific or location-specific and may not be transferable to different engineering standards. For instance, vulnerability curves developed for power transmission towers in South America (Reinoso et al., 2020) may not be applicable to towers designed under French standards. Nonetheless, curves developed for other regions or designs can sometimes be used illustratively when local information is unavailable. This practice serves to demonstrate the methodological approach, but it introduces substantial uncertainty, or, more precisely, a largely unquantified error, which limits the resulting study to a purely theoretical exercise.

### 2.4. Need for a Unified, Generalizable Vulnerability Modeling Framework for Adaptation to Climate Change

The reading of the literature, with our point of view, reveals three major gaps:

1. Lack of a generalizable vulnerability model across hazards, system types, system levels, and design characteristics. Existing models cover isolated hazard-system combinations and rarely form part of a larger, consistent vulnerability modeling architecture linking intensity of the climate related hazard and the physical damage rate of the system.
2. Insufficient availability of physical vulnerability curves. Particularly for infrastructures that may face climate-related hazards.
3. While climate modeling frameworks are starting towards this end, for the moment, there are no established standardized intensity-based methods for adaptation planning, especially methods relying on quantifiable Physical Damage Rate. Organizations responsible for infrastructure design, maintenance, and adaptation (including public agencies, network operators, insurers, and re-insurers) therefore lack quantitative physical damage rate-based information needed to evaluate long-term performance under evolving climate extremes and to set adaptation targets or investments priorities.

### 2.5. Vulnerability of Non-Infrastructure Physical Systems

In addition to infrastructure systems, other physical systems such as human physiological systems, natural systems and agricultural systems are also vulnerable to changing climate extremes.

Extreme temperatures cause mortality, health deterioration and loss working capacity (Poumadère et al., 2005; Stott et al., 2004; Swynghedauw et al., 2012). Water-related systems are affected by climate change (Edjossan-Sossou et al., 2023; Haddeland et al., 2014). Agricultural systems – including maize, rice, wheat, and soybean- can have yield damage under extreme temperatures, hydrological stress, extreme precipitation, wind extremes as explained in Feng et al., (2023); Li et al., (2025); Monteleone et al., (2023). Although this paper focuses on physical infrastructures, the general vulnerability modeling approach remains applicable to a wider set of physical systems exposed to climate extremes. In practice, climate adaptation challenges extend beyond infrastructures and concern human, natural, and agricultural systems. Thus, a generalizable vulnerability model should ideally be usable across these different physical systems whenever adaptation decisions require quantitative intensity-damage information.

### 2.6. Vulnerability of Infrastructure Systems

For infrastructures specifically, numerous studies across energy production, power transmission, roads, bridges, railways, and buildings reveal increasing exposure and potential damage under climate extremes (Koks et al., 2019; Verschuur et al., 2024). Wind energy structures can have some damage because of extreme wind gust intensities (Jaimes et al., 2020; Martín Del Campo et al., 2021). Hydropower and nuclear facilities face functional risks related to droughts and extreme temperatures (Parey et al., 2023). Extreme-value statistical modelling is important for energy infrastructures because it allows anticipating very low-probability, high intensity events, which cannot be captured using empirical data alone since the number of years in the observational record is limited  (Dutfoy et al., 2014; Parey, 2007). Power transmission infrastructures can be damaged by strong winds, as seen in the 1999 French storms (RTE, 2016). Vulnerability curves of power towers are provided by Reinoso et al. (2020) . Watson and Etemadi (2020) studies the wind damage to the transmission and power generating systems. Road networks face damage from extreme heat, clay shrink-swell, and extreme precipitation (Chang and Hossain, 2024; Ighil Ameur, 2023; Mulholland and Feyen, 2021; Sias et al., 2025; Underwood et al., 2017; Wang et al., 2024). Railway systems face hazards such as rail buckling under and catenary stress under extreme temperatures, damage under heavy precipitation (Ochsner et al., 2024; Zhang and Lin, 2025). Building systems face vulnerabilities to floods, landslides, wildfires, hail and extreme winds (Emanuel, 2011; Khanduri and Morrow, 2003; Maria Papathoma-Köhle et al., 2022; Papathoma-Köhle et al., 2019, 2015; Scawthorn et al., 2006). Collectively, these studies confirm that multiple infrastructure types are exposed to multiple hazards, and that damage is likely to increase under extreme intensities change.

### 2.7. Summary of the Research Gap

Given the change in climate extremes, infrastructures require adaptation to maintain acceptable performance. Quantitative vulnerability modeling is essential for identifying which infrastructures, subsystems, or components require adaptation, estimating the expected physical damage rate under different intensities, and determining the adaptation actions necessary to reduce physical vulnerability. Adaptation requires not only qualitative assessments but also precise quantification of expected damage. Without explicit physical damage rate-based information for each system, location, and hazard type, it becomes extremely challenging to prioritize adaptation actions, allocate resources, or design long-term adaptation strategies at the scale of assets, subsystems, infrastructures, or networks.

Despite this need, no existing framework fully and systematically links hazard intensity, system characteristics, and resulting physical damage across all hazards, system types, system levels, and design contexts. Framework such as CLIMADA (Aznar-Siguan and Bresch, 2019; Bresch and Aznar-Siguan, 2021; CLIMADA-project, 2026), HAZUS (Scawthorn et al., 2006; Vickery et al., 2006), and the multi-hazard vulnerability compilation paper from Nirandjan et al., (2024), contribute important steps by harmonizing portions of the vulnerability information that exist. However, they do not yet form unified and complete physical vulnerability model applicable across all systems and all climate-related hazards, containing all the needed information to do so. In this sense, the vulnerability curves available in literature, although heterogeneous, represent the best current basis from which a more general vulnerability modeling approach can be conceptualized. These scattered models collectively outline the direction toward a more systematic vulnerability framework, but many modeling gaps remain.
In this paper, we do not claim to fill these gaps, nor to provide operational outputs. The paper focuses on methodology and how it can be used in principle. We propose a theoretical and general formulation of physical-system vulnerability modeling based on vulnerability curves, consistent with what the literature suggests as the long-term direction for vulnerability modeling. Studies such as Nirandjan et al. (2024) come closest to assembling such a general model, by collecting vulnerability curves across multiple systems and hazards. Our contribution is therefore not to provide a complete operational vulnerability model, but rather to illustrate how such a general framework could be used for adaptation.

We do so through a purely theoretical use case, restricted to a very specific system (power transmission towers), for a single hazard (extreme wind gusts), and for a particular geographical domain (France). Based on strong assumptions, this use case shows in principle how a general vulnerability-curve-based framework could help estimate physical damage under given intensities, and how adaptation actions could be conceptualized by adapting the system and thus the associated vulnerability model. This does not constitute a final model, but rather an illustration of how existing vulnerability information, even if heterogeneous and imperfect, can be used within a general formulation to support vulnerability-based climate adaptation of physical systems.

# 3. Concepts

### 3.1. Physical Vulnerability and Systems Levels

Physical vulnerability modeling quantifies the physical damage rate experienced by a system when exposed to a climate-related hazard intensity. The input is a climate intensity $i$ that becomes a hazard intensity whenever the expected physical damage rate, $PDR$, is strictly positive. A system $s$ stands for an ensemble of infrastructure assets, an infrastructure asset, a subsystem, or a component, depending on the level at which vulnerability information is defined; system characteristics, design assumptions and the type of hazard are embedded in the parameter set $\theta(h, s)$.

### 3.2. Physical Damage Rate

The Physical Damage Rate, $PDR$, is a central output in the present work. $PDR$ is defined on $[0,1]$, with 0 indicating no damage and 1 indicating complete physical damage for a given system $s$ under intensity $i$ of hazard type $h$. Because it directly quantifies physical damage, $PDR$ is the relevant quantity for physical damage adaptation planning and long-term asset management.

### 3.3. Vulnerability Curves

Vulnerability curves (damage functions) link hazard intensity to expected physical damage rate for a given system and hazard. They provide intensity-conditioned damage information that can be used at the asset scale and aggregated to ensemble of assets. Figure 1 presents a theoretical vulnerability curve where physical damage rate increases with intensity.

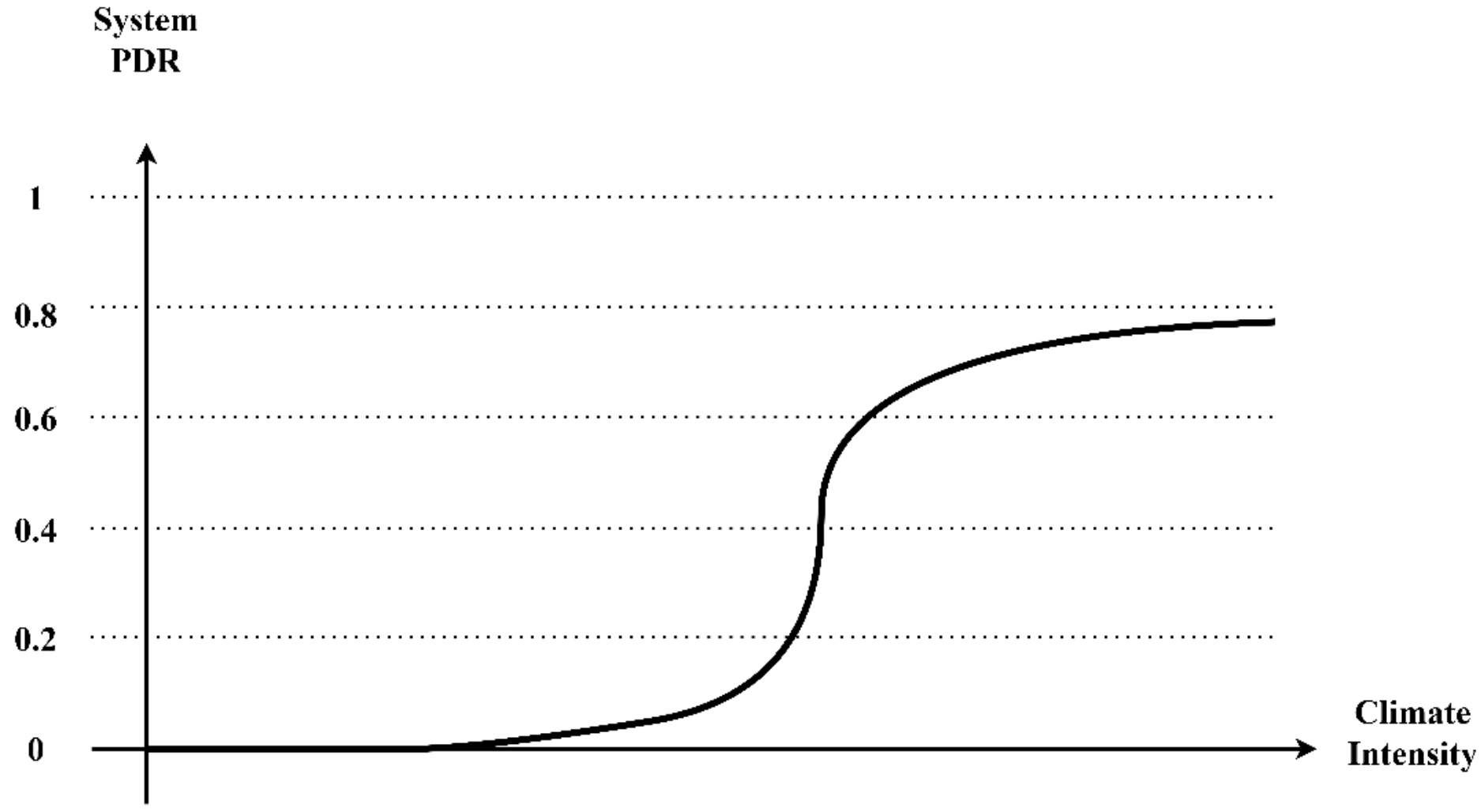

Figure 1. Theoretical Vulnerability Curve, where the Physical Damage Rate increases with climate intensity.

### 3.4. Fragility Curves

Fragility curves quantify the probability of failure as a function of intensity, in contrast to vulnerability curves that quantify physical damage rate of a given system. Fragility curves are essential for functional risk assessment and interdependent systems, but they do not provide $PDR$. Figure 2 shows a theoretical fragility curve where failure probability increases with intensity.

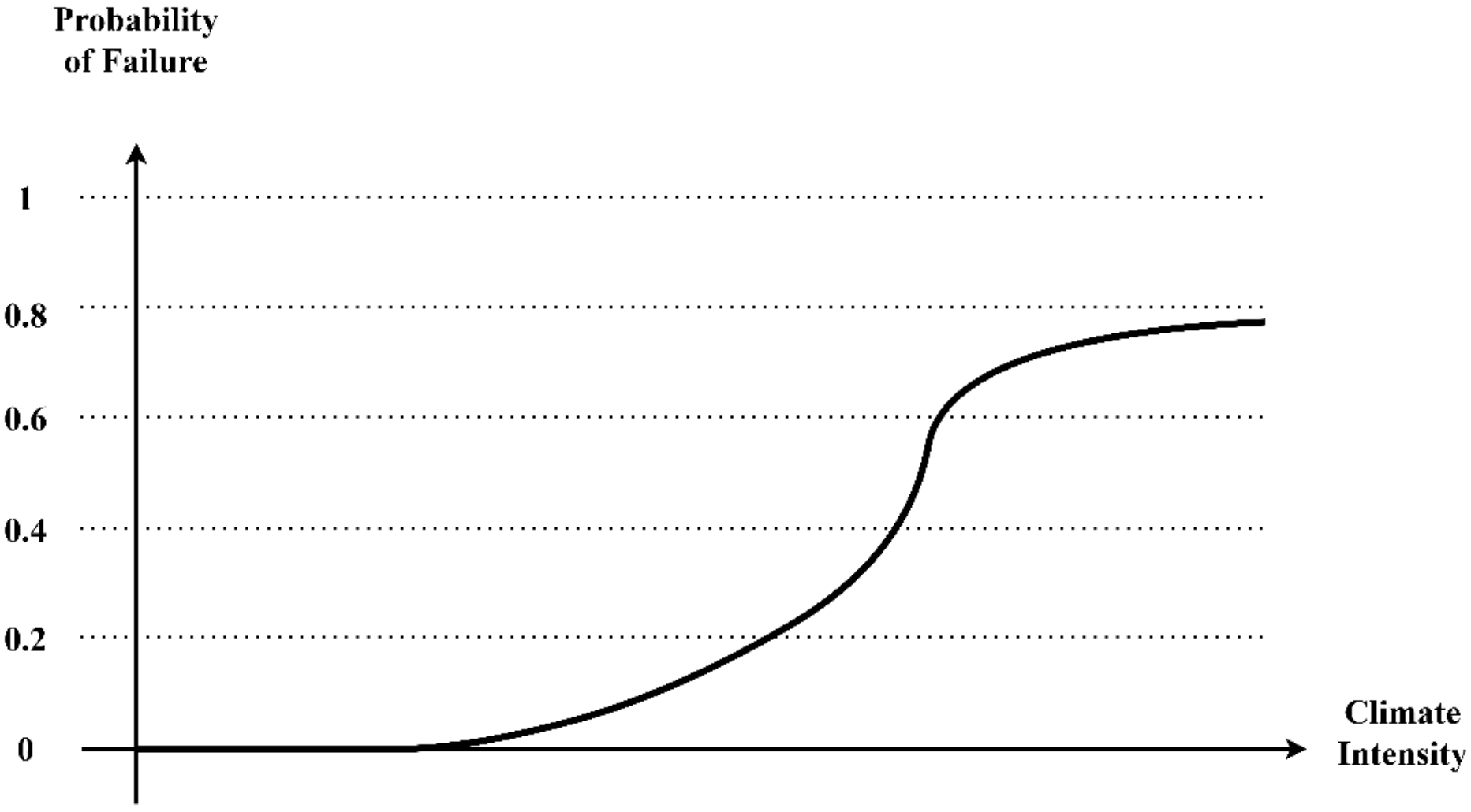

Figure 2. Theoretical Fragility Curve, the probability of failure of the system increases with climate intensity.

### 3.5. Vulnerability Indicators

Indicator-based approaches describe intrinsic system characteristics (often system-specific, design specific, location-specific), but they do not explicitly provide an intensity-damage relationship and therefore do not provide a specific physical damage rate given a specific intensity. They are useful when no curve exists for a given system and remain complementary to intensity-damage-based approaches, but are less suited for large-scale, intensity-based adaptation decisions.

# 4. The Vulnerability Model

### 4.1. Vulnerability Model

The general physical vulnerability model we formulate in this work is:

$$f_{\theta(h,s)}(i) = PDR_{s,i}$$

Where $h$ is the hazard type, $s$ the system (at a given system level), $i$ the climate intensity, $PDR_{s,i}$ the expected physical damage rate, and $\theta(h,s)$ the system and hazard specific parameters. This formulation represents the ideal state of knowledge for linking intensity and damage.

### 4.2. Adaptation Through Vulnerability Reduction

Adaptation by reducing vulnerability corresponds to a rightward shift of the vulnerability curve on the intensity axis. Quantitatively, for the same intensity $i$, the expected $PDR$ after adaptation is inferior to the initial state:

$$\forall i, \qquad f_{\theta(h,s)}^{adapted}(i) < f_{\theta(h,s)}^{initial}(i)$$

In practice, decisions would not be made for all intensities but for a specific return intensity $i_T$ , associated with a given return period $T$.

$$For\ a\ given\ i_T, \qquad f_{\theta(h,s)}^{adapted}(i_T) < f_{\theta(h,s)}^{initial}(i_T)$$

In the context of physical vulnerability reduction, adaptation is represented by modifying the vulnerability curve of a system so that, for the same climate intensity in input, the expected $PDR$ decreases. This corresponds to shifting the vulnerability curve to the right on the intensity axis, meaning that a higher intensity is required to reach the same level of physical damage rate. This conceptual representation is shown in Figure 3, which illustrates three possible system designs with different intrinsic vulnerabilities (so with different system characteristics) and how adaptation corresponds to selecting, or transitioning toward, a less vulnerable design.

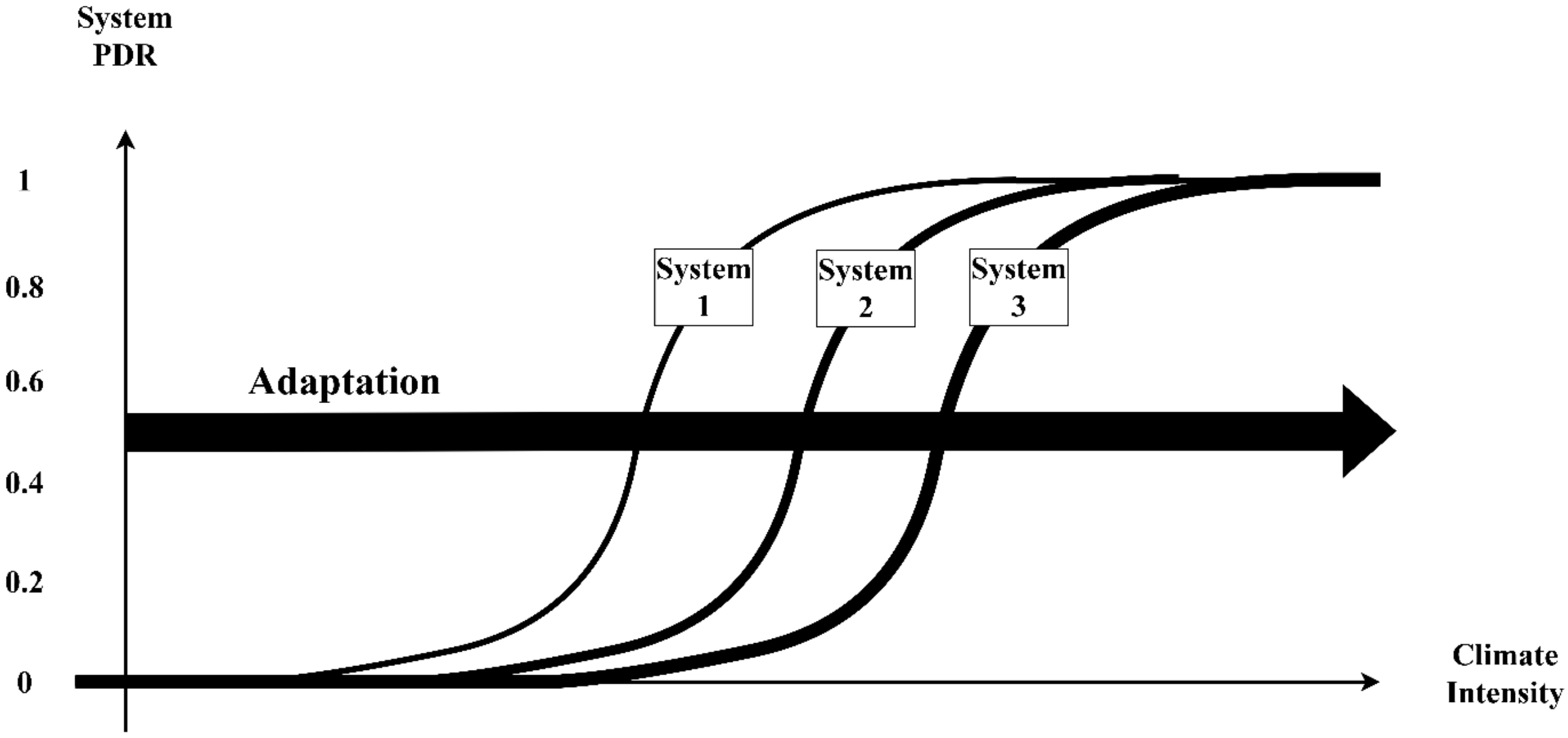

Figure 3. Adaptation via shifting vulnerability curves.

In addition to this qualitative representation, vulnerability-based adaptation decisions often focus on a specific intensity of interest, typically associated with a return period or a design value. Figure 4 presents several vulnerability curves evaluated at a fixed intensity threshold $i_T$, often corresponding to a specific return period. This allows comparison of different system designs at the same climatic intensity and highlights how the physical damage rate varies depending on the intrinsic vulnerability of each system. Such figures support decisions where the aim is to select or design a system that exhibits a lower $PDR$ at the intensity $i_T$.

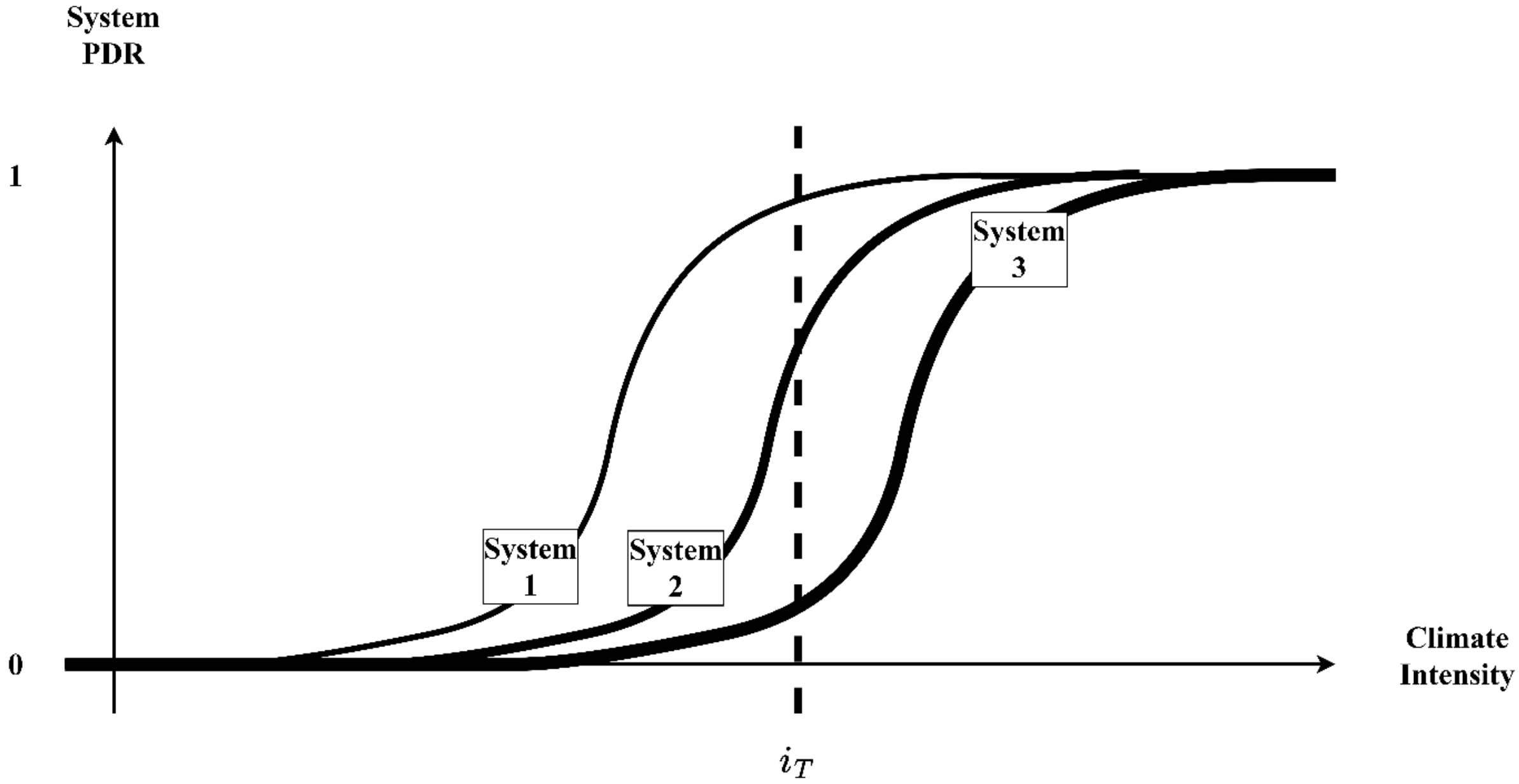

Figure 4. Vulnerability curves for different systems at a given intensity threshold $i_T$.

### 4.3. Fragility-Curve-Based Adaptation

Although the focus here is physical damage rate, adaptation may also be considered on the fragility side as reducing the failure probability for a given intensity. Fragility-based adaptation is conceptually parallel to vulnerability-based adaptation: improving the design or characteristics of a system results in a rightward shift of the fragility curve, meaning that a higher intensity is required to reach the same probability of failure. This representation is shown in Figure 5, where different system designs show different intrinsic probabilities of failure as intensity increases.

Fragility-based adaptation is particularly relevant for networked or independent systems, in which failure of one component can propagate and stop the functionality of the entire network. In such contexts, fragility curves provide essential information about the likelihood of system level or cascading failure to a whole network, which cannot be inferred from physical damage rates alone. This perspective belongs to the broader field of functional vulnerability and resilience modeling, which includes network independence and functional performance under extreme events.

However, while fragility-based approaches are highly valuable for assessing functional performance and resilience, they are not developed further in this paper, because the present work focuses explicitly on physical vulnerability and physical-damage-based adaptation. The aim in this paper is to quantify physical damage rate under hazard intensities and to illustrate how to reduce intrinsic vulnerability. And thus, how to reduce projected physical damage rate for given scenarios implying unseen extreme intensities. Fragility-driven adaptation is therefore acknowledged as an important and complementary perspective, but it lies outside the scope of the current methodology. In fact, fragility-based adaptation could form an entire study on its own, since fragility functions express the probability of failure of a system as a function of hazard intensity.

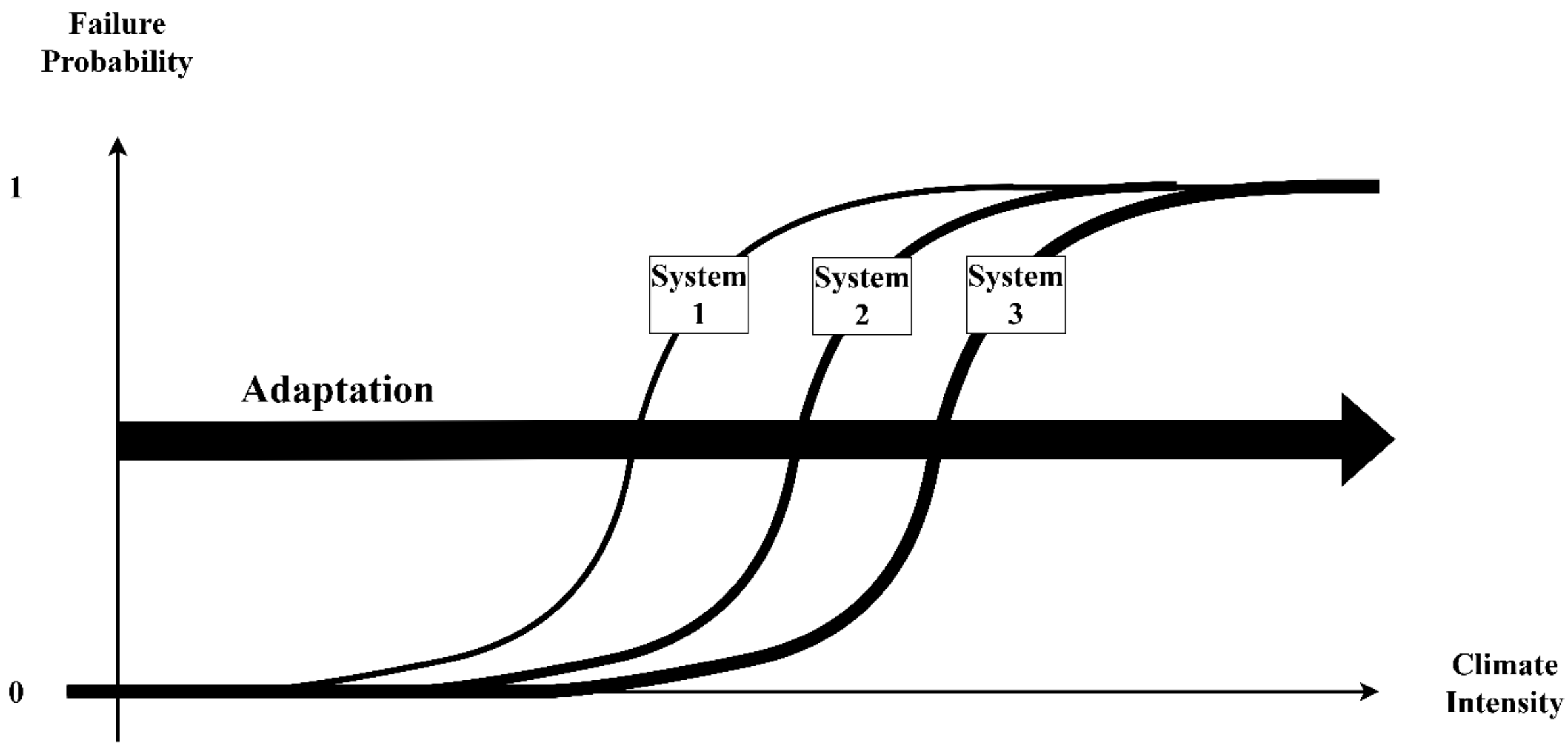

Figure 5. Fragility curves illustrating adaptation actions directed by reduction of probability of failure.

# 5. Methodology

### 5.1. Required information

The practical implementation of the vulnerability-based adaptation approach requires three elements.

(i) Vulnerability curves for the considered system type and hazard, describing the relationship between hazard intensity and the expected physical damage rate for the specific system level and design.

(ii) The geographical distribution of the systems, so that each asset or system location can be assigned to a climate intensity value extracted from the climate dataset.

(iii) Climate intensities over selected time periods, expressed as climate intensities corresponding to return periods of interest, and defined on a spatial grid covering the study area.

These three prerequisites correspond directly to those used in the theoretical case study developed later, where vulnerability curves are taken from existing literature, system locations are derived from spatial data, and climate intensities come from gridded wind-gust datasets.

### 5.2. Step-by-Step Physical Vulnerability Based Adaptation

Step 1. Assign climate intensities to system locations. Retrieve $i$ for each asset using nearest-grid or downscaling process to obtain the climate intensity at the right location. Then empirical or statistical extreme value analysis is performed to obtain return intensity values for the right location for the right return period.

Step 2. Compute physical damage rate at each location. Apply the relevant vulnerability curve to obtain physical damage rate for each system.

Ideal Step 2. We need to explain some ideas here in case of generalization of vulnerability modelling. The step 2 can be performed for many system types and hazards. In that sense, a general vulnerability model would be particularly useful: instead of computing damages along a single infrastructure type or subsystem, one could assemble system-specific and hazard-specific vulnerability curves into a broader and more comprehensive vulnerability model. Such a model would make it possible to represent, within a single structure, the different system levels, designs, and hazard types already identified in the existing literature. It would also allow the same methodological steps to be applied consistently across multiple systems and hazards, each using its own appropriate vulnerability curve. This perspective directly answers the fragmentation issues highlighted in the research gap section: existing curves are heterogeneous and context-specific but together they form the foundations of a larger vulnerability framework in which each curve contributes as a precise and well-defined piece of the overall model. However, in the present theoretical use case, we apply the computation to one infrastructure type and one hazard.

Step 3. Map physical damage rate. Produce physical damage rate maps for a given return period for a given system or ensemble of systems.

Step 4. Evaluate adaptation. Compare the physical damage rate obtained for each system, for each system location to the acceptable damage levels defined for the study and for a given system (for different systems acceptable damage level may be different e.g. a nuclear power plant does not have the same criticality as a country road). For each system type, and each location if the acceptable levels are exceeded, adaptation is required.
Thus, a database can be produced to show which system type is needed to respect the physical damage rate constraint for a given intensity.

Step 5. Outputs. A database can be produced to indicate, for a given intensity, which system design (or which vulnerability curve) is required to satisfy the physical damage rate constraint.

We stop here the adaptation method in this study. But of course, for practical adaptation, a lot more steps are required, among others: quantification of adaptation finance and planning construction work to adapt infrastructures to climate change.

### 5.3. Handling Missing or Non-Local Curves

A key limitation is that vulnerability curves are not available for all assets, hazards, or system levels. This is why, to date, a precise general vulnerability model (valid for all hazards, all systems, all subsystems, all locations, and all designs) does not exist. The basis for such a general model nevertheless exists in the literature, but for many system-type, hazard-type pairs with a given design, no precise curve is currently available.

To obtain a precise vulnerability model that quantifies damage, there is no alternative to investing in: (i) data collection to obtain climate intensities, system characteristics, and damage observations; (ii) modeling to link given intensity to a given damage for a specific system; (iii) time and funding to perform the above.
Even with such an investment, uncertainties will remain. If such an investment is not possible, indicator-based methods can help estimate the relative intrinsic vulnerability of one asset compared to an ensemble of assets. However, with the indicator-based quantification, it will be difficult to determine the physical damage rate. Similarly, an existing vulnerability curve from another context (e.g., not exactly the same design as the study's asset) may be used with caution to obtain an order of magnitude of damage, provided the associated assumptions are explicit.

### 5.4. Methodological Limitations

To the best of our knowledge, the method is limited by the state of existing knowledge: incomplete availability of precise vulnerability curves; spatial resolution of intensity data; challenges to get non-stationarity of extremes using historical or projected climate intensity data. While these limitations will decrease as climate risk research and industry grow, they do not prevent methodological development but need to be explained explicitly. These methodological limitations imply use cases hypotheses which at their turn imply uncertainty and errors in the results of use cases.

The next section applies this methodology to a theoretical case with strong hypotheses: power transmission towers (spatial distribution hypothesized from open data), wind-gust intensities over selected periods (we select the nearest intensity values of a low-resolution grid, and do not perform extreme value statistics), and vulnerability curves from the literature for different designs. The case is not meant to be operational; it illustrates how the general vulnerability model can be used to compute physical damage rate and to discuss adaptation via vulnerability reduction.

## 6. Use Cases: Set Up & Results

We developed a purely theoretical case study based on a set of hypotheses that are intentionally unrealistic or inappropriate for the actual system under consideration. These hypotheses are made deliberately in order to illustrate the methodology, and they imply that none of the results presented in this section can be directly applied to real-world infrastructure systems. We assume that the vulnerability of the system is known, although in reality it is not. We also assume that the vulnerability curves perfectly correspond to the system studied, even though the actual technical characteristics of the system are unknown in this study. As a consequence, we adopt vulnerability curves from similar systems located in other countries, and we apply them to French power-transmission towers. This assumption is clearly imperfect in real-world context, but for the purpose of this methodological illustration, we treat these curves as if they were representative of the studies system. In the same spirit, we imagine that the spatial distribution of the infrastructure system is known. We imagine also that the spatial distribution of extreme climate intensities is sufficient and we take the nearest climate intensity value near the infrastructure without downscaling. We assume that each system is associated with available vulnerability curve, even though such information is often missing or incomplete in practice.

Under these assumptions, we compute the expected physical damage rate for each system at each location, using the corresponding extreme intensity. When the resulting damage exceeds a defined acceptable threshold, we then reduce the vulnerability, by choosing a less vulnerable curve or design, so that the damage remains below the threshold. In this theoretical framework, adaptation is therefore driven by identifying the vulnerability curve that provides the desired physical performance level for the hazard intensity considered.

### 6.1. System: Set Up

The system considered consists of power transmission towers derived from open-source data. The spatial distribution is obtained from OpenStreetMap by extracting 225kV and 400kV transmission lines and assuming a tower every 425m along these power lines. This produces a simplified and hypothetical spatial representation of the network, shown in Figure 6. Because the data are not provided by the system operator and the spacing assumption of the power tower is generic, the resulting distribution does not correspond to the actual infrastructure: it is used solely for illustrative purposes. Although the use case focuses on power transmission towers, the same methodological steps could be applied to other types of physical infrastructures exposed to climate hazards (e.g., wind turbines or residential buildings) provided that appropriate vulnerability curves are available.

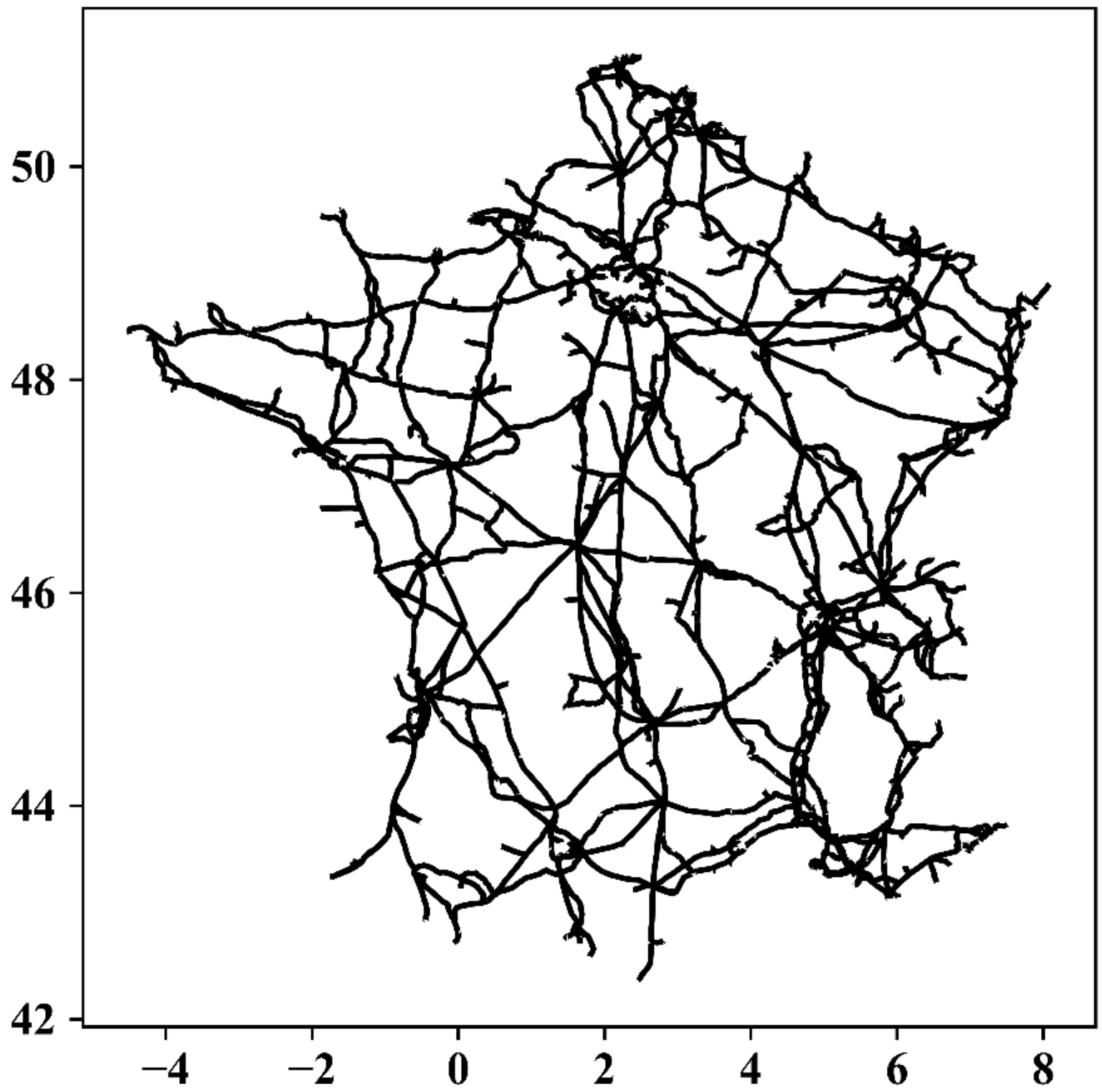

Figure 6. Spatial repartition of the system. This system represents 225kV and 400kV power towers derived from overhead power lines of 225kV and 400kV. Data are sourced from Open Street Map.

**6.2. Climate Intensity: Set Up**

Wind-gust intensities are taken from the ERA5 reanalysis dataset, extracted for two periods: 1940-1982 and 1982-2024. We split the historical period 1940-2024 in two different datasets in order to be able to see if there if return intensities are the same or not. No statistical extreme value modeling is performed within the use case, instead empirical return intensities are taken on the 42 years periods, for each point. No downscaling on the intensity data is performed. The resulting intensity maps for both periods are shown in Figure 7. These intensities represent return-period wind-gust estimates empirically derived from the dataset, acknowledging that their spatial resolution and representativeness are limited.

The climate intensity set up is very minimalist on purpose, the goal is to focus on the vulnerability modeling, and vulnerability-based adaptation methodologies. The

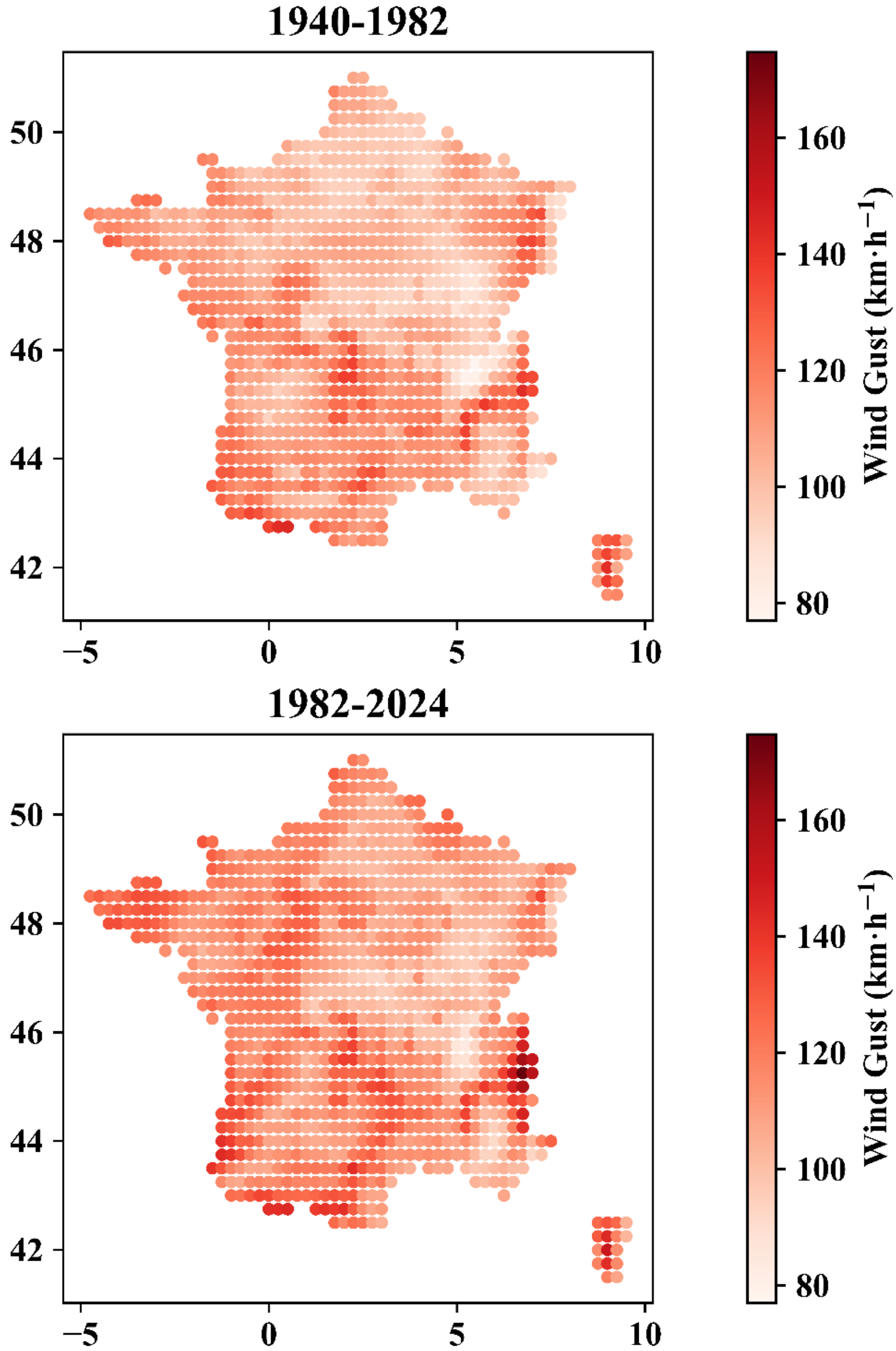

Figure 7. Wind gust intensities for the periods 1940-1982 and 1982-2024 used as climate intensity inputs in the use case.

### 6.3. Vulnerability Model: Set Up

The vulnerability model used in this use case links the climate-intensity input to the physical damage rate that the system may experience. In general, a vulnerability model may consist of a single vulnerability curve or a set of vulnerability curves, each representing a specific system type or a different structural design. In principle, such a model could be generalizable and made applicable to any system and any climate hazard. Achieving such a fully general vulnerability model, however, would require detailed information on all relevant climate intensities, on all system characteristics, and on the exact physical damage associated with each system-hazard pair. Such comprehensive information is not available.

Among existing studies, the work of Nirandjan et al., (2024) is particularly relevant. It brings together vulnerability information for several system types and several hazards, and, for some system types, it even provides multiple vulnerability curves, each corresponding to a distinct structural design. This does not constitute a complete vulnerability model covering all hazards, systems, system levels, and design types, and it should not be regarded as the best possible approximation of such a general model. Rather, it is the closest existing study to this direction, as it compiles and organizes the vulnerability information currently available in the literature, thereby opening a path toward a general formulation introduced earlier in this paper.
In this use case, we use the vulnerability curves assembled by Nirandjan et al. (2024), which themselves originate from Reinoso et al., (2020). These curves were initially developed for power-transmission towers corresponding to Mexican design standards. For the purpose of this theoretical illustration, we make the explicit and strong assumption, acknowledged as incorrect, that these curves correspond to the French power towers considered here. This assumption is one of the reasons why the use case is purely theoretical, but it allows us to demonstrate how the methodology operates when vulnerability curves are available.
The set of curves used in the use case is shown in Figure 8. Each curve represents the physical damage rate as a function of wind-gust intensity and corresponds to a different tower design; the name of the curve is characterized by a design wind speed.

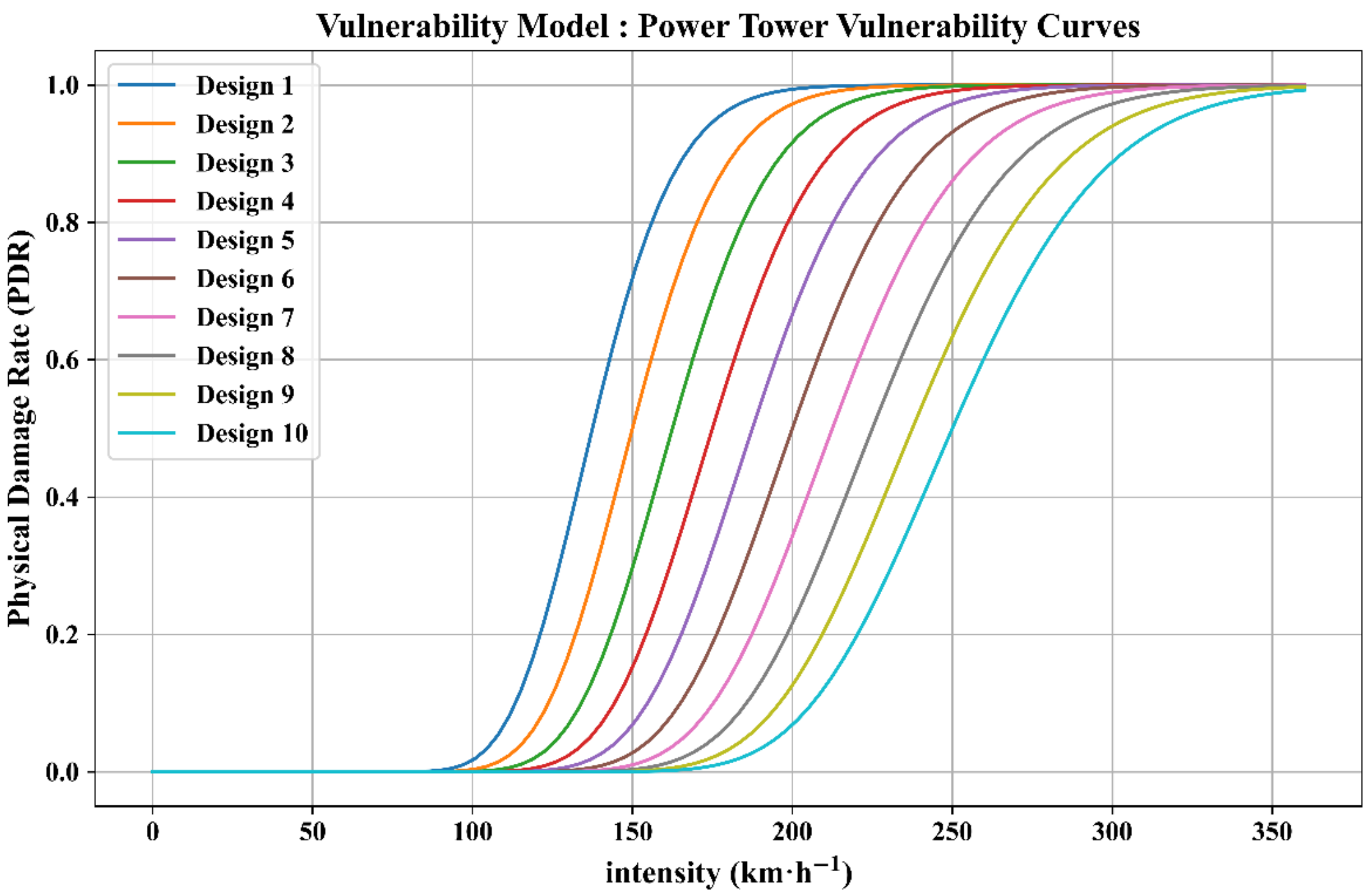

Figure 8. Vulnerability Model: Power Tower Vulnerability Curves
This figure presents a set of vulnerability curves for power towers facing wind-gust intensities. The horizontal axis shows wind intensity (km.h-1), ranging from 0 to above 350 km.h-1, while the vertical axis shows the Physical Damage Rate, $PDR$, from 0 (no damage) to 1 (complete damage). Each curve corresponds to a specific tower design and labelled from the most vulnerable to wind speed (Design 1) to the least vulnerable (Design 10). The curves originate from Nirandjan et al. (2024) which compiled power-tower vulnerability data originally developed by Reinoso et al. (2020).

### 6.4. Computation of the Physical Damage Rate, $PDR$

The computation of the physical damage rate requires three elements that have been defined in the preceding subsections: (i) the climate intensities, here represented by wind-gust empirical return intensities derived from the two empirical 42-year periods, (ii) the system, represented here by hypothetically distributed power-transmission towers over the entire French territory, and (iii) the vulnerability model, which consists of several vulnerability curves corresponding to different tower designs.

While each of these inputs is simplified or hypothetical – the spatial distribution does not reflect the actual French network, the tower characteristics are unknown to us, and the vulnerability curves originate from South American systems (Nirandjan et al., 2024b; Reinoso et al., 2020) – we proceed as if the assumptions were valid. The unique objective of this exercise is to illustrate the methodology, not to produce operational conclusions. For the purpose of the demonstration, we further assume that all towers share the same baseline design: the most vulnerable curve available in the set (Design 1). This is a strong and acknowledged simplification, real tower networks often exhibit heterogeneous design across the territory, but it allows the method to be presented clearly without the additional complexity of design heterogeneity.

With these elements in place, the projected physical damage rate for each tower location is obtained as follows. Given a map of intensity for a given return period over the territory, the damage at each tower site results from applying the chosen vulnerability curve to the local intensity. Because the intensity grid and the tower locations do not perfectly coincide spatially, the link is established through a nearest-neighbor assignment: each tower receives the intensity of the nearest grid point. The computation can be carried out in two equivalent orders. Either by first computing the damage at every grid point and then assigning the resulting damage value to each tower by nearest neighbor proximity. Or by first assigning the nearest grid point intensity to each tower and then applying the vulnerability curve directly at the tower sites. Both sequences yield the same result, since nearest-neighbor assignment is used in both cases. Through these steps, an expected physical damage rate value is obtained for each tower at each location, producing a spatial map of physical damage across the system.

The resulting spatial distributions for the two empirical periods are shown in Figure 9. As previously explained, these values do not represent real expected damages, given the non-local origin of the vulnerability curves and the hypothetical nature of the system. Their purpose is purely illustrative: they demonstrate how the methodology transforms climate-intensity data into a spatial distribution of physical damage rate using vulnerability curves specific to different designs of a given system type, here: power towers.

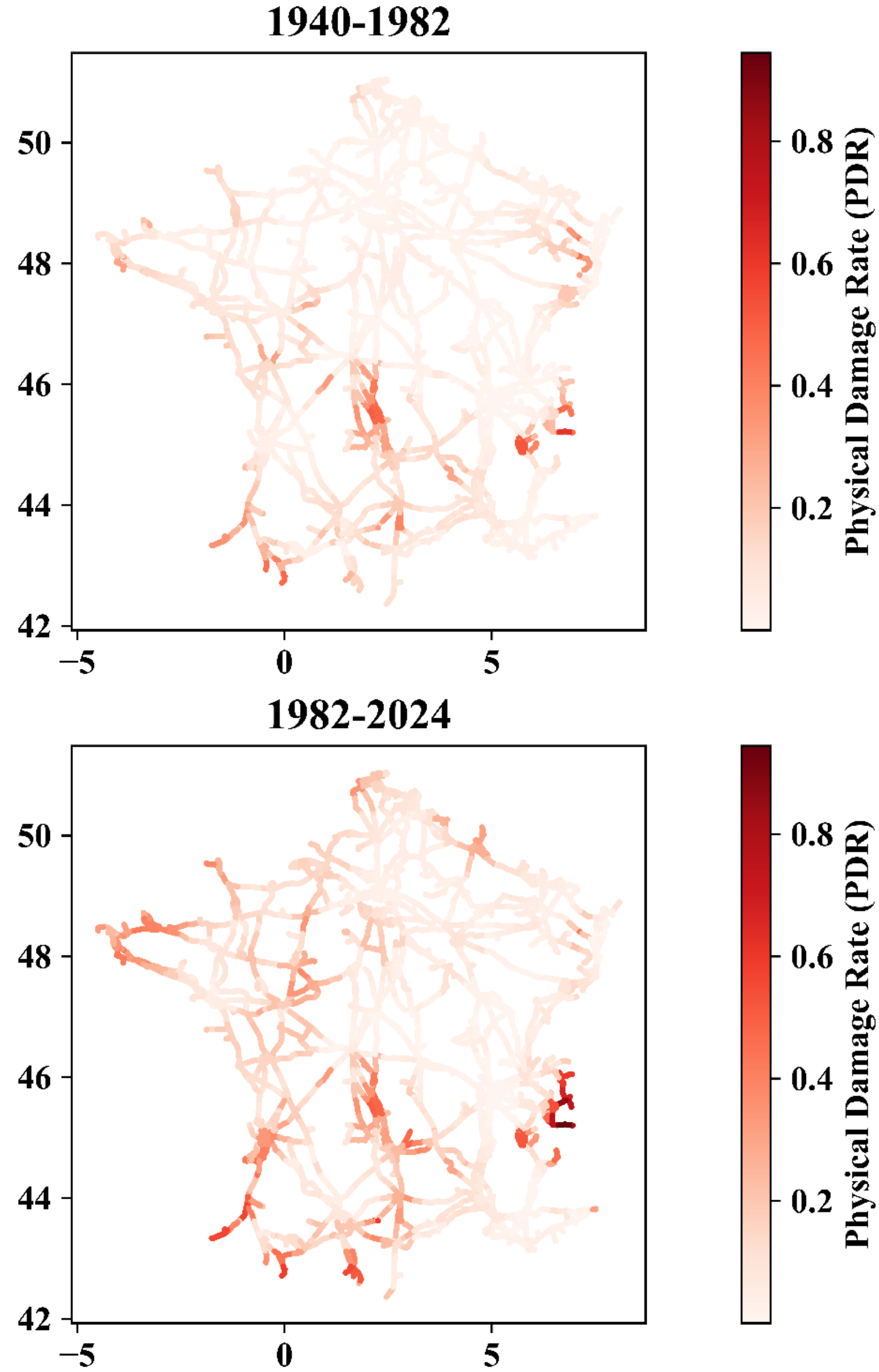

Figure 9. Physical Damage Rate spatial distribution for a hypothetical power tower ensemble based on an initial design (Design 1) on the empirical periods 1940-1982 and 1982-2024.

Once the physical damage rate has been computed for each tower location, the next step consists in comparing it to an acceptable damage threshold defined by the asset manager or the study (here we take the threshold of 0.1 of physical damage rate). When the threshold is exceeded at a given location, adaptation is required. In practice, each design corresponds to a different vulnerability curve, characterized by a specific design wind speed. By evaluating the physical damage rate produced by each available curve at each location, it becomes possible to identify the minimal design that keeps the expected damage below the acceptable threshold. This selection process directly indicates the level of system vulnerability required under a given climate condition. In the context of continuing climate change, the same methodology applied to climate projections would reveal how the required design level evolves over time, thereby informing adaptation planning for the asset portfolio.

The resulting spatial distributions of adapted design choices for the two historical periods are shown in Figure 10. The maps highlight where a less vulnerable design would be necessary to comply with the physical damage rate constraint under evolving wind-gust extreme intensities, illustrating the adaptation of physical infrastructures to changing extreme wind conditions. While this use case is applied to historical periods, the identical methodology can be carried out using climate projections to support forward looking adaptation decisions.

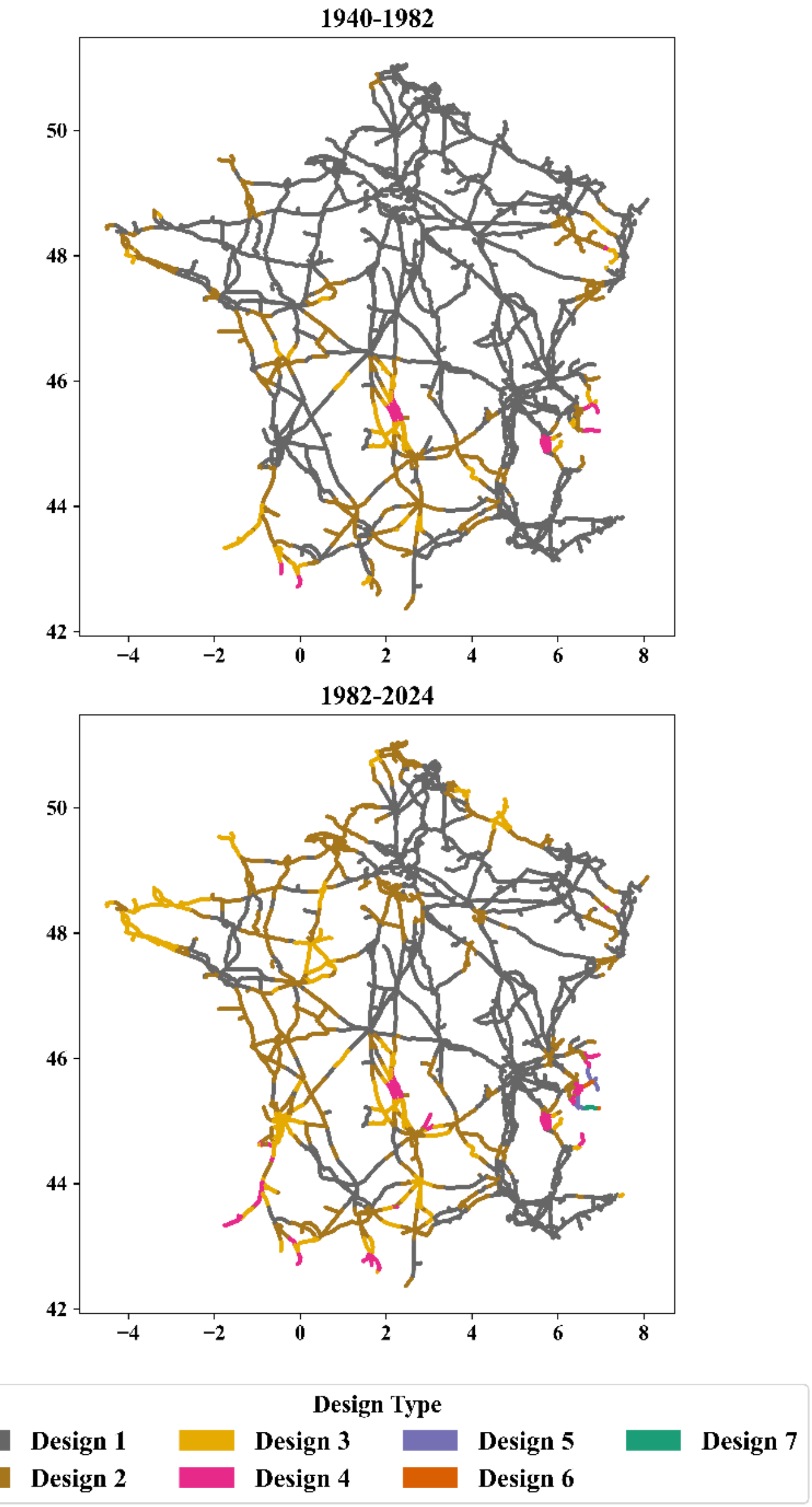

Figure 10. Adapted Design Choices satisfying the physical damage rate constraint for the two time periods 1940-1982 and 1982-2024.

# 7. Discussion & Limitations

The vulnerability model presented in this paper links the climate-intensity input to the physical damage rate that a system may experience. A full general vulnerability model, applicable to any system and any climate data, would in principle require information on all relevant climate intensities, on all system characteristics, and on the exact physical damage associated with each system-climate data pair. Such precise information is not available, and this absence constitutes a central constraint within which the present methodology operates.

The methodology is constrained by critical assumptions at each step. Assigning climate intensities to system locations (Step 1) requires both precise knowledge of system geolocation and high-resolution climate data that is corresponding to the system, making this step highly sensitive to climate and system data quality and granularity. The application of vulnerability curves and PDR computation (Step 2) assumes the existence of accurate, system-specific curves and their availability across diverse designs. In practice, these assumptions are often unrealistic: where system-specific vulnerability curves are unavailable, practitioners must rely on curves from similar systems, introducing unquantified errors into PDR estimates. While Steps 3-5 (mapping, adaptation evaluation, design outputs) do not introduce major additional hypotheses, their validity remains dependent on the validity of precedent steps. Any imprecision in Step 1 and 2 propagates through these steps without being corrected. Steps 3 to 5 are therefore limited by the errors accumulated through the entire methodology.

# 8. Conclusion

This paper has established the theoretical and methodological foundations for modeling the physical vulnerability of systems to climate-related hazards for adaptation of physical systems, especially infrastructures to withstand a given physical damage rate. The general vulnerability formulation, $f_{\theta(h,s)}(i) = PDR_{s,i}$ , provides a systematic and generalizable framework for linking climate intensity, system characteristics, and expected physical damage rate. The methodology requires three elements: vulnerability curves describing the intensity-damage relationship for specific system types and designs, the geographical distribution of the systems, and climate intensities which can be expressed as return-period values over the study area. The step-by-step procedure enables the computation of physical damage rates across spatial domains, the identification of locations where acceptable damage thresholds are exceeded, and the selection of appropriate system designs to meet performance constraints.

The theoretical use case on power transmission towers facing wind-gust hazards across France has demonstrated how this framework operates in practice. The case has shown how climate-intensity data can be transformed into spatial distributions of physical damage rate through the application of vulnerability curves, and how adaptation actions can be derived by selecting vulnerability curves corresponding to less vulnerable designs. However, the use case has also made explicit that it is purely theoretical, relying on strong and acknowledged as incorrect assumptions regarding the transferability of vulnerability curves from one engineering context to another, the homogeneity of system designs, the treatment of empirical intensities without extreme value statistics, and the use of modified open-source spatial data that does not reflect actual infrastructure distribution.
Future research should aim at building specialized and generalizable vulnerability models, spanning multiple hazards, system types, system levels, and design characteristics. This requires addressing the scarcity of physical vulnerability curves, which remains a structural limitation of the field, by producing and consolidating vulnerability information across systems and hazards. The present work stops, even from a theoretical point of view, at the modification of the system based on its vulnerability curve. A natural extension would be to go one step further and integrate financial data into the use case, so that physical damage rates can be translated in economic terms and adaptation options compared on a cost basis.

# 9. CRediT authorship contribution statement

Matthieu Dutel: Conceptualization, Methodology & Models, Software, Validation, Formal Analysis, Investigation, Data Curation, Writing-Original Draft, Visualization.

# 10. Acknowledgement

Matthieu Dutel thanks Sixense and its clients, engineers, asset managers and experts; thanks LGI team for research advice; thanks Jean Meunier Pion for scientific listening; thanks Anne Barros, Adam Abdin, Didier Soto for supervision during CIFRE partnership. Part of this paper was developed in the context of a CIFRE doctoral partnership with CentraleSupélec and Resallience by Sixense Engineering, VINCI Group from 1.03.2023 to 28.02.2026. The end of this paper is self-funded by Matthieu Dutel from 1.03.2026 to 30.07.2026.